# Skewness of citation impact data and covariates of citation distributions:

# A large-scale empirical analysis based on Web of Science data

Lutz Bornmann* & Loet Leydesdorff**

* Division for Science and Innovation Studies

Administrative Headquarters of the Max Planck Society

Hofgartenstr. 8,

80539 Munich, Germany.

E-mail: bornmann@gv.mpg.de

** Amsterdam School of Communication Research (ASCoR)

University of Amsterdam

PO Box 15793

1001 NG Amsterdam, The Netherlands

E-mail: loet@leydesdorff.net

**Abstract**

Using percentile shares, one can visualize and analyze the skewness in bibliometric data across disciplines and over time. The resulting figures can be intuitively interpreted and are more suitable for detailed analysis of the effects of independent and control variables on distributions than regression analysis. We show this by using percentile shares to analyze so-called "factors influencing citation impact" (FICs; e.g., the impact factor of the publishing journal) across year and disciplines. All articles (n= 2,961,789) covered by WoS in 1990 (n= 637,301), 2000 (n= 919,485), and 2010 (n= 1,405,003) are used. In 2010, nearly half of the citation impact is accounted for by the 10% most-frequently cited papers; the skewness is largest in the humanities (68.5% in the top-10% layer) and lowest in agricultural sciences (40.6%). The comparison of the effects of the different FICs (the number of cited references, number of authors, number of pages, and JIF) on citation impact shows that JIF has indeed the strongest correlations with the citation scores. However, the correlation between FICs and citation impact is lower, if citations are normalized instead of using raw citation counts.

# 1 Introduction

van Raan (2014) listed the skewness of citation data as one of several methodological problems in citation analysis. The skewness of bibliometric data has been a topic in this field since its beginnings in the 1920s. The issue is associated with the "laws" of Alfred Lotka, Samuel Bradford, and George Zipf: "the concentration of items on a relatively small stratum of sources" (de Bellis, 2009, p. xxiv). Since then a large number of papers have appeared demonstrating the skewness of citation data. Seglen (1992), for example, argued that "50% of the citations and the most cited half of the articles account for nearly 90% of the citations" (p. 628). He concluded that citation distributions follow approximately an inverse power-law distribution (the number of citations larger than $x$ is proportional to -log($x$) (Katz, 2000). Albarrán and Ruiz-Castillo (2011) showed empirically that the "existence of a power law cannot be rejected in ALL SCIENCES taken together as well as in 17 of 22 fields whose articles represent 74.7% of the total" (p. 48). Using a replication and scale invariant technique – the Characteristic Scores and Scales (CSS) (Glänzel, 2011) – the results of Albarrán, Crespo, Ortuño, and Ruiz-Castillo (2011) show that citation distributions are highly skewed: „the mean is 20 points above the median, while 9–10% of all articles in the upper tail account for about 44% of all citations" (p. 385).

In this study, we analyze the skewness of citation impact data in six major disciplines (natural sciences, engineering and technology, medical and health sciences, agricultural sciences, social sciences, and humanities) based on all articles in Web of Science (WoS) published in 1990, 2000, and 2010. First, we use percentile shares – a recently introduced visualization and analysis technique – to quantify the proportions of total citation impact that go into different groups (e.g., the 10% most-frequently-cited papers). Percentile shares can be intuitively and appealingly interpreted and are especially suitable "for the detailed analysis of distributional changes" (Jann, 2016, p. 3).

In a next step, we use percentile shares to analyze covariates of the citation distributions. Journal Impact Factors (JIF) are often used as proxies for the citation impact of papers published in the respective journals. Are JIFs indeed a factor influencing citation impact? We show the advantages of using percentile shares in the case of a number of co-variates of citation scores indicated in the literature as "factors influencing citation impact" (e.g., the number of authors, see Bornmann & Daniel, 2008). We compare the association of JIFs as co-variates with citation scores at the level of individual papers with other FICs mentioned in the literature, such as number of co-authors, the numbers of pages, and the number of cited references. How much does each co-variate enhance the likelihood of being cited in the top-10% layer of citation scores? Finally, we address the question of whether normalization of the citation scores increases or reduces this chance.

Let us note that the analysis is correlational. One cannot conclude that FICs influence citation impact since a third factor such as the quality of the paper may be involved (Bornmann & Leydesdorff, 2015). If a paper is of high quality, e.g., it may attract large numbers of citations and be published in a journal with a high impact. The relation between the JIF and the citation impact is then spurious. From this perspective, the terminology "factors influencing citation" is perhaps unfortunate.

## 2    Factors influencing citation counts: a short review of the literature

In the following, only a short review focusing on recent and overview studies is provided. Recently, Tahamtan, Safipour Afshar, and Ahamdzadeh (2016) published a comprehensive review of factors influencing citation counts (FICs). The results of many studies question the usefulness of citation counts for measuring research impact or of using citation counts as a proxy for research quality. Overviews of FICs listed in tabular forms can be found in Onodera and Yoshikane (2014) and Didegah and Thelwall (2013). Both tables emphasize the Journal Impact Factor (JIF, Garfield, 2006) as an important factor in receiving citations.

In a recent study, Didegah and Thelwall (2014) investigated a range of factors which may be associated with the citation counts of social-science papers. The authors conclude that "journal and [cited] reference characteristics, and particularly journal and reference impact, are the main extrinsic properties of articles that associate with their future citation impact in the social sciences. Journal and reference internationality can also help with the prediction of future citation impact. Research collaboration, and particularly individual and institutional collaboration, can help to predict citation counts for articles but international collaboration alone is not important, unless it is with a high impact nation. Paper length, abstract length and abstract readability are also significant determinants of citation counts, but not all make a substantial difference. In the world top institutions, articles with more readable abstracts (i.e., easier to read) were less cited but in the social sciences more readable abstracts are more cited" (Didegah & Thelwall, 2014, pp. 173-174).

Robson and Mousquès (2016) focused on papers in environmental modelling published since 2005 and studied a range of FICs which were quantified or classified. The results of the study reveal that "papers with no differential equations received more citations. The topic of the paper, number of authors and publication venue were also significant. Ten other factors, some of which have been found significant in other studies, were also considered, but most added little to the predictive power of the models. Collectively, all factors predicted 16-29% of the variation in citation counts, with the remaining variance (the majority) presumably attributable to important subjective factors such as paper quality, clarity and timeliness" (Robson & Mousquès, 2016, p.

94). Onodera and Yoshikane (2014) studied samples of papers in six selected fields (condensed matter physics, inorganic and nuclear chemistry, electric and electronic engineering, biochemistry and molecular biology, physiology, and gastroenterology) and tried to reveal some general patterns. "Some generality across the fields was found with regard to the selected predicting factors and the degree of significance of these predictors. The Price index [the proportion of references within 3 and 5 years (de Solla Price, 1963)] was the strongest predictor of citations, and number of references was the next. The effects of number of authors and authors' achievement measures were rather weak" (Onodera & Yoshikane, 2014, p. 739).

For the field of "information science & library science", Yu, Yu, Li, and Wang (2014) used stepwise regression to produce a model predicting citation counts. The authors included a range of possible FICs and concluded that they can predict – with relative good accuracy – citation impact using a citation window of five years.

# 3 Methods

### 3.1 Percentile shares and Gini coefficients

We build on our argument for using percentiles in bibliometric evaluations (Bornmann, Leydesdorff, & Mutz, 2013; Leydesdorff, Bornmann, Mutz, & Opthof, 2011). Here, percentiles are used to field- (and time-) normalize citation counts; a percentile is the percentage of papers with lower or higher citation impact – depending on the percentile formula. Hicks, Wouters, Waltman, de Rijcke, and Rafols (2015) posit that "the most robust normalization method is based on percentiles: each paper is weighted on the basis of the percentile to which it belongs in the citation distribution of its field (the top 1%, 10% or 20%, for example)" (p. 430).

This study introduces the statistical method of "percentile shares" developed by Jann (2016) in scientometrics. Whereas percentiles are frequently used in bibliometrics to count top-cited papers, percentile shares focus on the inequality of distributions. For example, in the case of the proportion of income earned by the top-1% or top-10% of inhabitants, "(p)ercentile shares quantify the proportions of total outcome (e.g. of total income) that go to different groups defined in terms of their relative ranks in the distribution. They thus have an intuitive and appealing interpretation and can be used for detailed analysis of distributional changes" (Jann, 2016, p. 3).

The percentile shares were calculated with the "pshare" command in Stata (StataCorp., 2015) which estimates percentile shares of an outcome variable (here: citation impact data) from individual level data (here: papers). A detailed description of the method can be found in Jann (2016). For example, linear interpolation is applied in regions where the empirical distribution function is flat. Additionally, percentile shares allow determining "the relative ranks of the population members using an alternative outcome variable" (Jann, 2016, p. 10). This is used in the present study to analyze how citation impact is distributed across different FICs groups (e.g. the bottom 50%, mid-40%, and top-10% of papers in terms of numbers of authors).

In addition to percentile shares, we report Gini coefficients. Gini coefficients are measures of inequality (Gini, 1997, 1909): a coefficient of 1 means that a single paper receives all citations and a coefficient of 0 means that the citations are equally distributed over the papers.

### 3.2 Dataset used

The bibliometric data were harvested from the in-house database developed and maintained by the Max Planck Digital Library (MPDL, Munich) in October 2016. The study is based on all papers in the WoS with the document type "article" (n= 2,961,789) published, respectively, in 1990 (n= 637,301), 2000 (n= 919,485), and 2010 (n= 1,405,003). We select these publication years in order to study changes in skewness of citation impact data and the relatedness of citation impact with FICs in the long run of two decades. We study the relationship between citation scores and (i) the number of cited references, (ii) the number of authors, (iii) the number of papers, and (iv) the JIF of the publishing journal at the level of individual papers.

Two FICs which are used in this study have missing values: 776,500 papers are published in journals without a JIF (in MPDL) and the number of pages is not provided in the case of 13,220 papers. The large numbers of missing values for the JIF is mostly due to the fact that the MPDL in-house database does not contain JIFs for 1990. Thus, the analyses using these two variables are based on reduced data sets.

Papers are categorized to six OECD classifications of disciplines: natural sciences, engineering and technology, medical and health sciences, agricultural sciences, social sciences, and humanities. These disciplinary classifications are aggregates of WoS subject categories clustering journals in the database. [1]

## 4 Results

### 4.1 The representation of skewness using percentile shares

Figure 1 shows percentile share densities of citation impact in the six different disciplinary categories. The figure is based on all articles from 1990, 2000, and 2010 and their citations during a three-year citation window (subsequent to the publication year). The restriction to a three-year citation window is justified by the fact that one can expect more citations for papers published a long time ago than for papers published recently. Thus, "the age of an article is to some extent controlled" (Neophytou, 2013). Glänzel and Schoepflin (1995) conclude that a three-year citation window is a good compromise between the fast obsolescence of some fields (e.g., nanotechnology) and the slow obsolescence of other fields (e.g., theoretical mathematics).

---

[1] http://ipscience-help.thomsonreuters.com/incitesLive/globalComparisonsGroup/globalComparisons/subjAreaSchemesGroup/oecd.html

The densities in the figure indicate how much impact each paper in a discipline receives (on average) in relation to the overall citation impact. The results are shown for three impact groups: papers belonging to the bottom-50%, mid-40% (that is, the 50 - 90% range), and the top-10% in terms of citations. Furthermore, the results are separately visualized for the natural sciences, engineering and technology, medical and health sciences, agricultural sciences, social sciences, and humanities. Figure 1 shows that papers in the top-10% group receive citation impact which is between five and eight times above the average impact. This highest value of approximately eight times can be observed in the case of the humanities; in all other disciplines the values are significantly lower. Although the results in the figure are in agreement with many other studies on citation impact distributions (see above), percentile shares visualize the skewness in different disciplines very clearly.

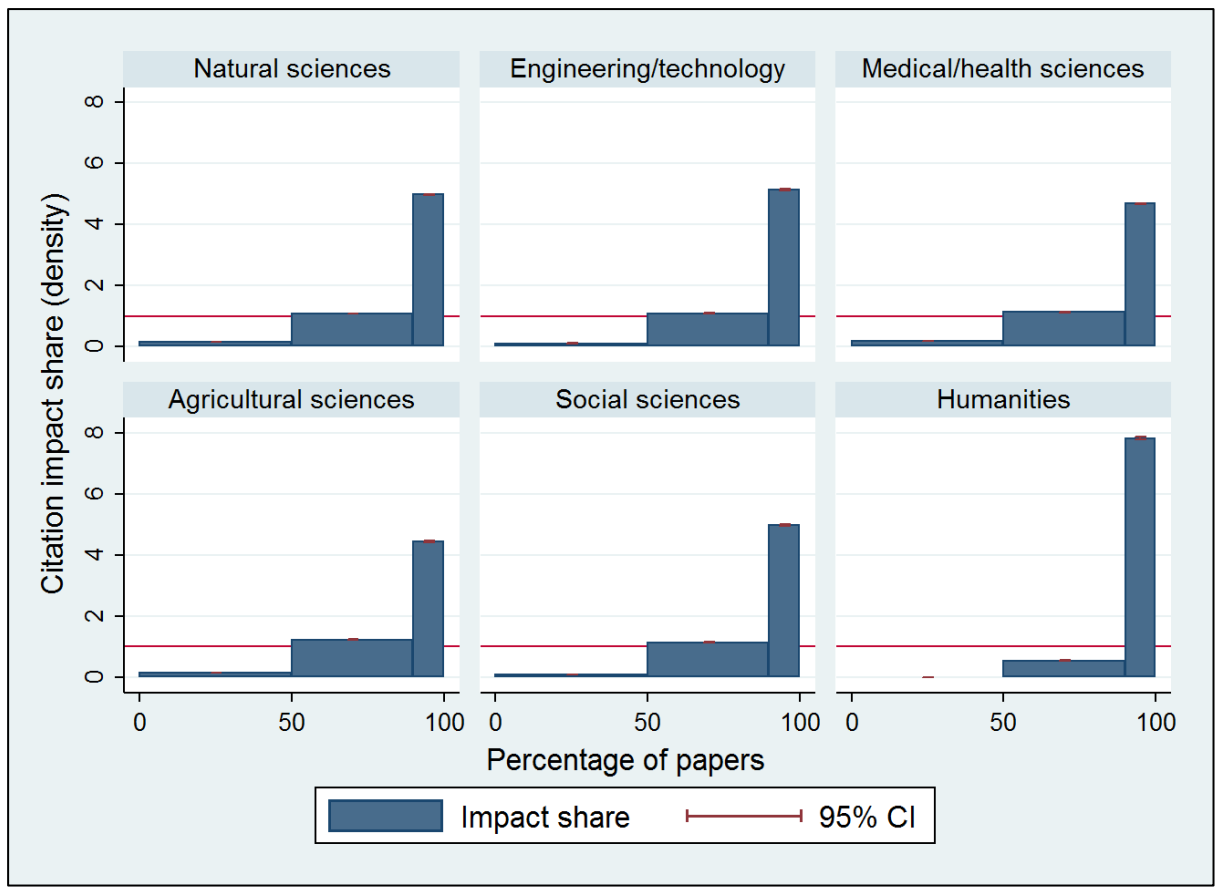

Figure 1. Percentile share densities of citation impact (bottom 50%, mid-40%, and top-10%) for a three year citation window (excluding the publication year) over papers published in different disciplines

We extend the analyses of citation distributions now by showing differences between the years 1990, 2000, and 2010 and by comparing the citation distributions with linked cited references distributions. Linked cited references are that subgroup of all cited references, which could be matched with a publication record in WoS. In contrast to the results in Figure 1 (where a fixed citation window is used), the citation window for each paper in the following figures is from publication year (e.g., 1990) up to the end of 2014.

Figure 2 (left side) shows the percentage of citation counts in different disciplines which fall upon the bottom-50% (blue segment), mid-40% (red segment), and top-10% (green segment) of papers. The percentages of citation counts in Figure 2 (left side) are compared with the percentages of linked cited references of the same papers (Figure 2, right side). We can see in Figure 2 (Times cited, 1990), for example, that the 10% of papers from 1990 with the most citation impact received 57.6% of the total impact (which all papers from 1990 have received). The bottom-50% of papers in terms of citation impact received less than 10% of total impact. This skew in the distribution of impact is visible for all disciplines. The citation impact distribution is most-highly skewed in the case of the humanities: the top-10% received 75.6% of the total citation impact (in 1990). Note that the Arts & Humanities Citation Index (A&HCI) itself is relatively stable in the period under study (Leydesdorff, Hammarfelt, & Salah, 2011).

The comparison of the publication years (1990, 2000, and 2010) shows that the share of impact accounting for the top-10% segment is decreasing which is especially visible for the social sciences (from 66.4% in the 1990 to 45.1% in 2010). Thus, the results point to a decreasing focus of citing authors on the top-cited papers. During these two decades (1990-2010), social scientists increasingly moved from a national orientation towards adopting international standards and thus became more similar to the other sciences (Digital Science, 2016; Merton, 1973). In addition to percentile shares, Figure 2 also shows Gini coefficients along the right axis: in agreement with the percentile share results, the Gini coefficients point to a decreasing focus on the top-cited papers.

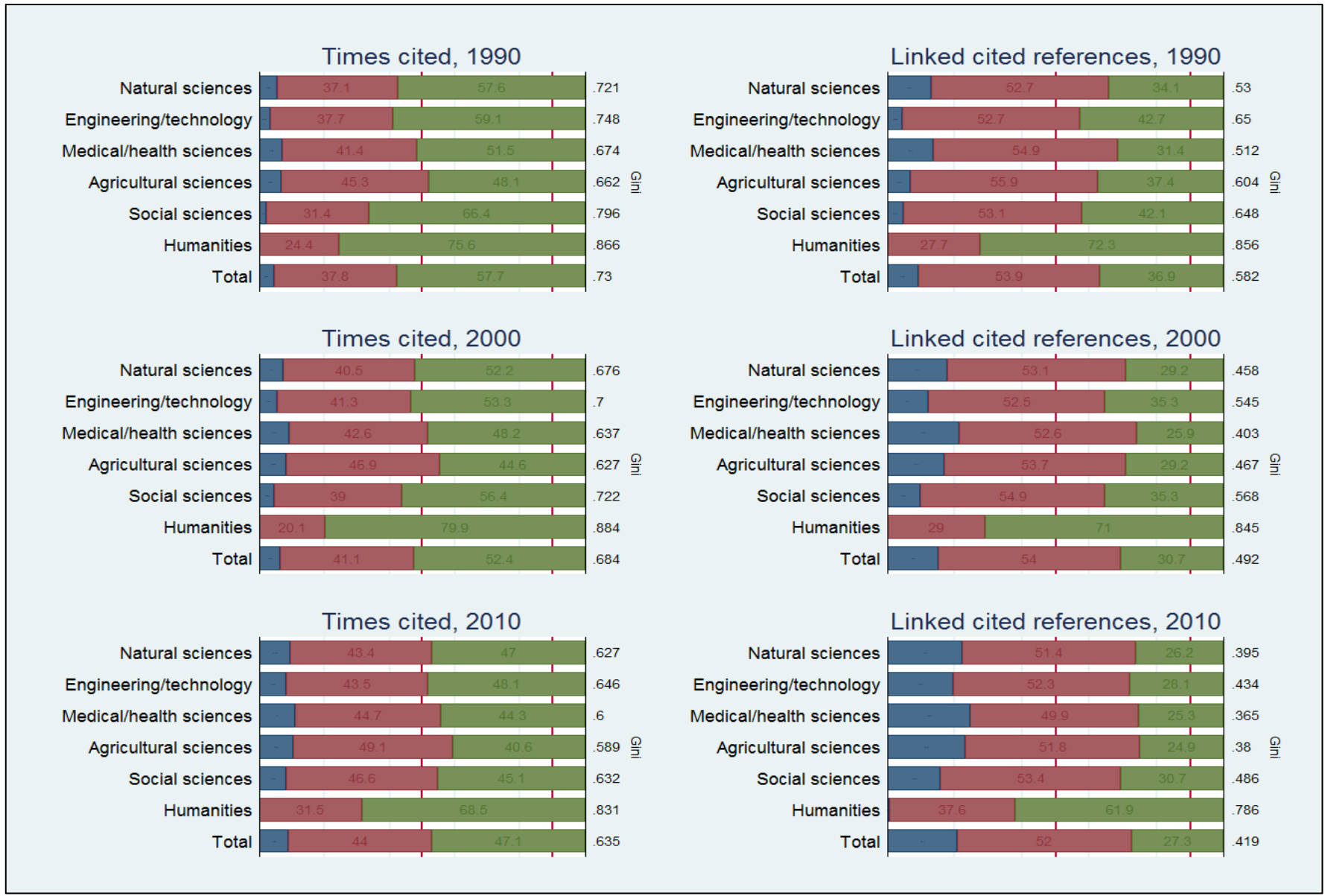

Figure 2. Percentage of citation impact and linked cited reference counts in different disciplines which fall upon the bottom 50% (blue segment), mid-40% (red segment), and top-10% (green segment) of papers (measured in terms of citations and linked cited references, respectively)

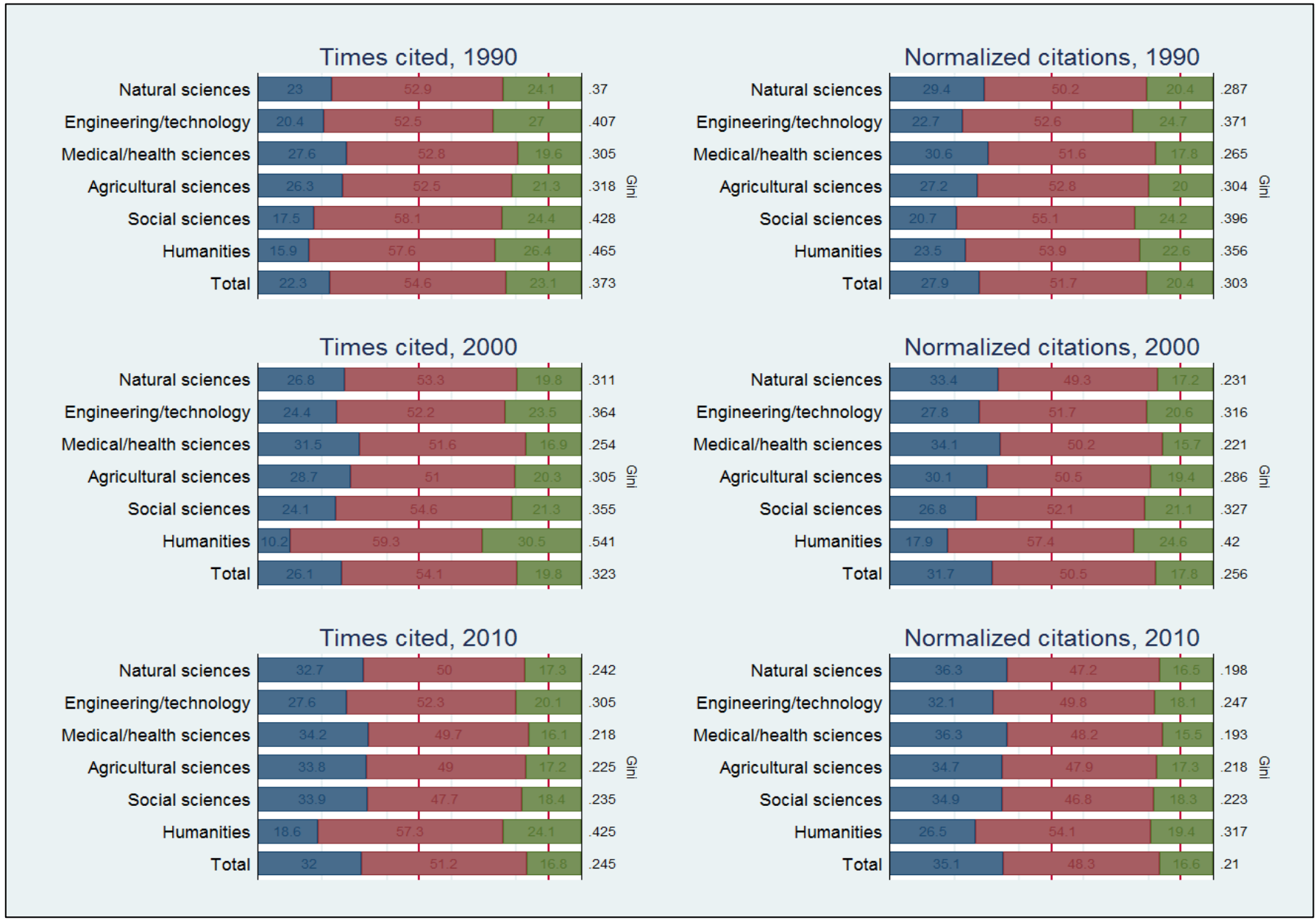

Figure 3. Percentage of citation impact which falls upon the bottom 50% (blue segment), mid-40% (red segment), and top-10% (green segment) of papers in terms of all (linked and not-linked) cited reference counts in different disciplines

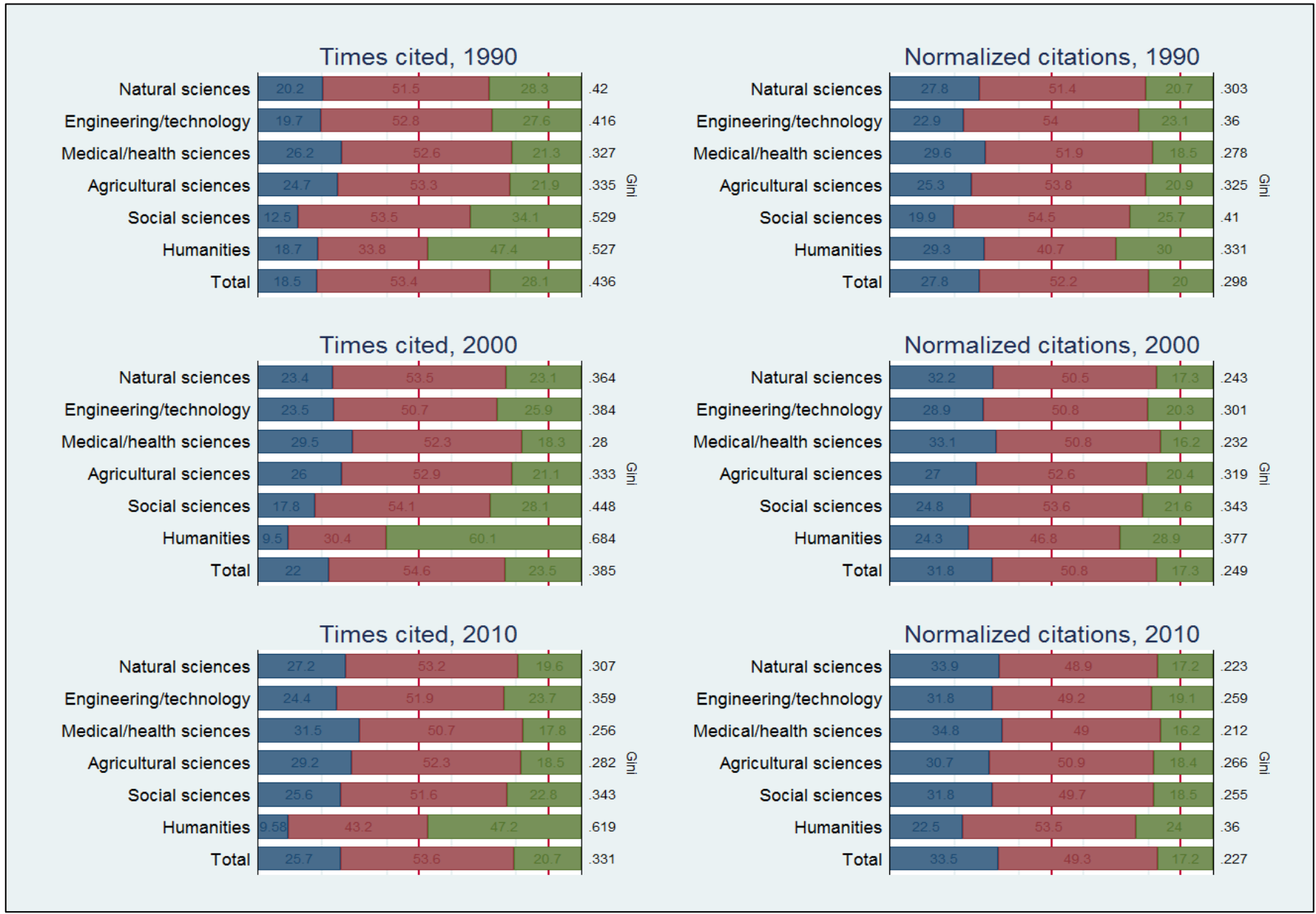

Figure 4. Percentage of citation impact which falls upon the bottom 50% (blue segment), mid-40% (red segment), and top-10% (green segment) of papers in terms of linked cited reference counts in different disciplines

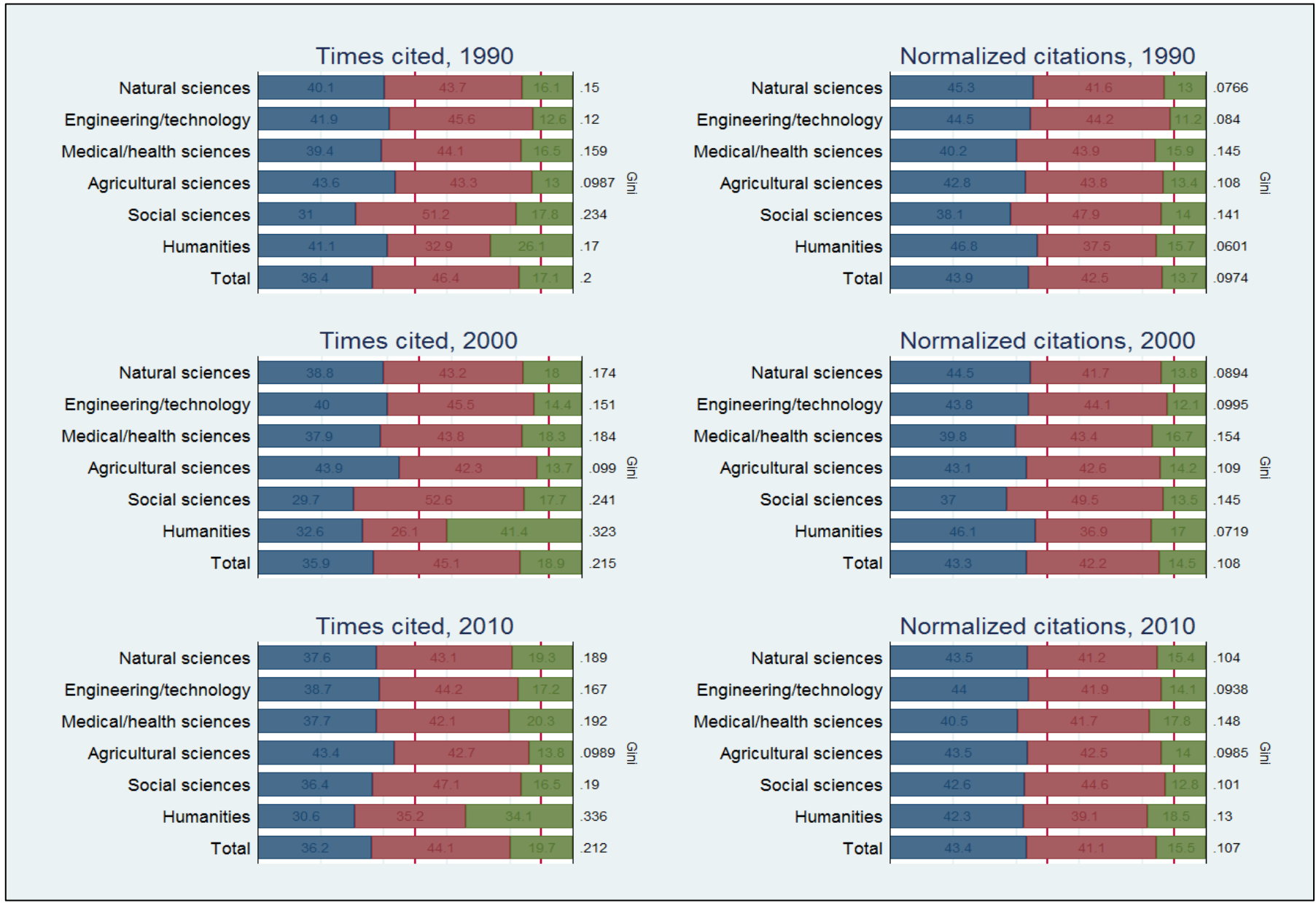

Figure 5. Percentage of citation impact which falls upon the bottom 50% (blue segment), mid-40% (red segment), and top-10% (green segment) of papers in terms of author counts in different disciplines

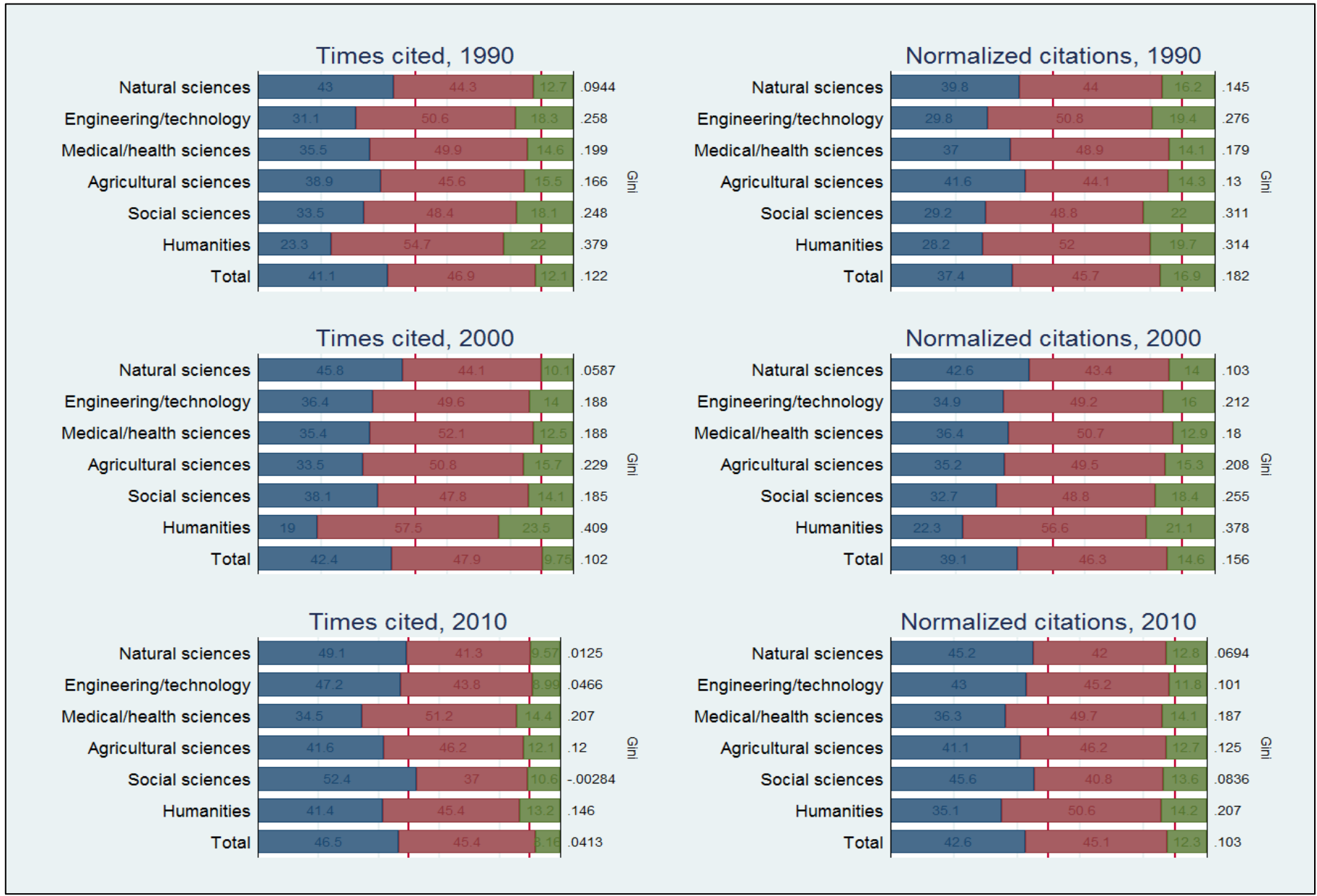

Figure 6. Percentage of citation impact which falls upon the bottom 50% (blue segment), mid-40% (red segment), and top-10% (green segment) of papers in terms of page numbers in different disciplines

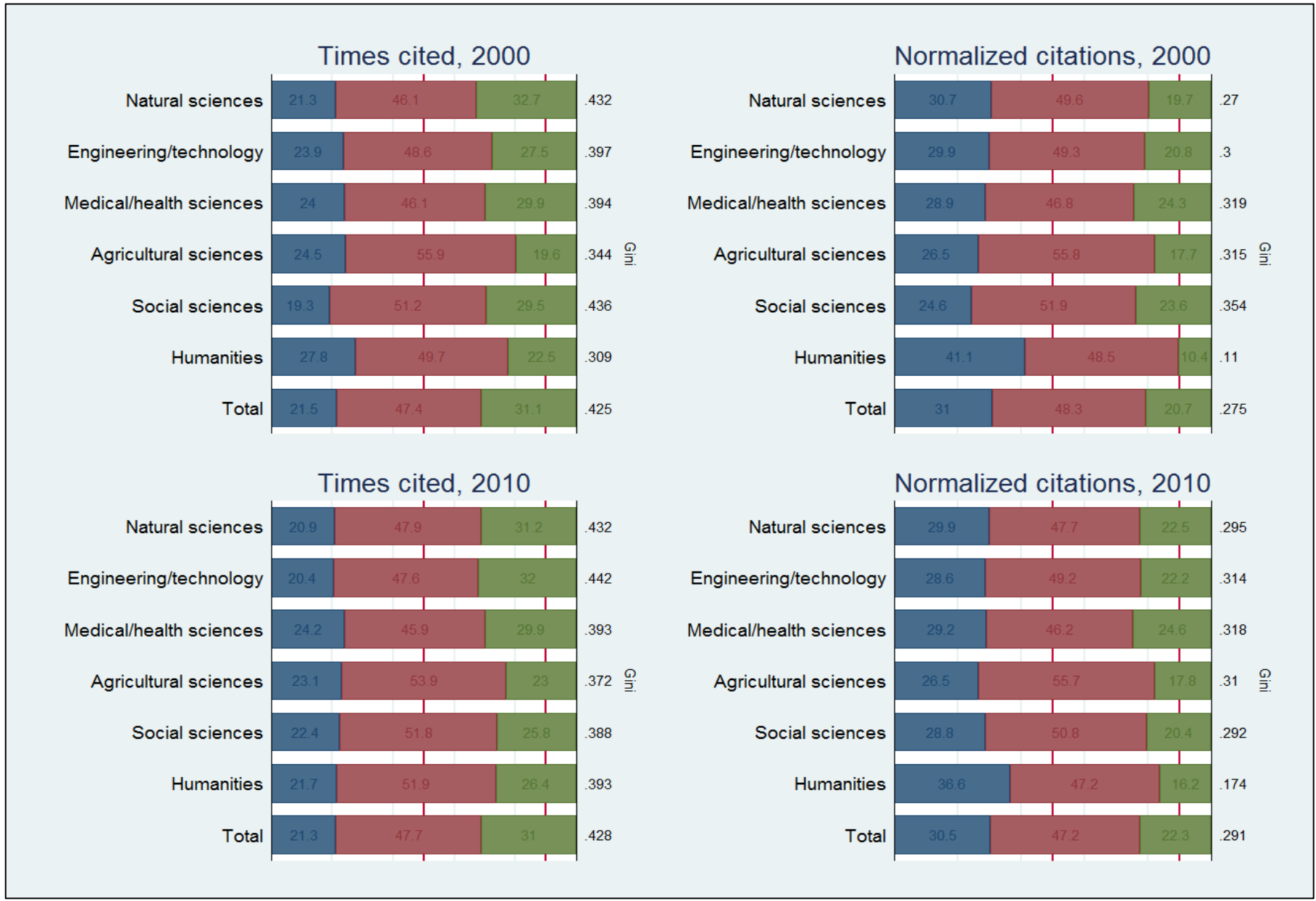

Figure 7. Percentage of citation impact which falls upon the bottom 50% (blue segment), mid-40% (red segment), and top-10% (green segment) of papers in terms of journal impact (measured by the journal impact factor, JIF) in different disciplines

The times cited information in the WoS database is derived from the cited references extracted from the citing papers: Citations are the number of matches between the cited references of citing papers and the bibliographic information of records in the database. Linked cited references are that part of all cited references which could be matched not only on the citing side, but also on the cited side. Because the domain is then delineated, one may expect similar distributions of linked cited references and citations: The bar charts on the right side of Figure 2 show the distributions of the number of linked cited references in the papers from 1990, 2000, and 2010. The distributions of the cited references show similar patterns as the distributions of the citations. However, the distributions are not as skewed as the citation counts distributions. This is also indicated by the lower Gini coefficients.

### 4.2 Percentile shares and factors influencing citations (FICs)

In this section, we extend the analysis of citation distributions using percentile shares by further considering FICs. Figure 3 to Figure 7 visualize the relationships between citation counts and each FIC under study. All figures have the same layout: The left side shows the results based on the times cited information from the WoS database; the right side shows the results for the same data normalized against the mean citation score in the reference set (MNCS, see Opthof & Leydesdorff, 2010). The MNCS-indicator is field and time normalized: each paper's citation impact is divided by the mean citation rate in the corresponding field (WoS subject category) of the publishing journal and the publication year (Bornmann & Marx, 2015). Although the mean citation rate should not be considered as the expected rate because of the skew in the distributions, the MNCS is used as a standard indicator in bibliometrics.

The results for the MNCS are additionally shown in the figures because it will be tested whether the normalization lowers the correlation between FICs and citations. For example, the number of authors is one of the FICs in this study. The mean number of authors is different in the subject categories within the disciplines. If the citation impact of the papers is normalized by the mean citation rates in the categories, the effect of the number of authors on the citation impact of the papers might be reduced. The reduction would be an important argument for the use of normalized citations for research evaluation purposes, since citation scores should be influenced as little as possible by factors which are extrinsic to the substantive quality of the papers themselves.

Figure 3 shows the percentage of citation impact which falls upon the bottom-50% (blue segment), mid-40% (red segment), and top-10% (green segment) of papers in terms of all (linked and not-linked) cited reference counts in different disciplines. For example, the results for "Times cited, 2010" (Total) reveal that the top-10% of papers in terms of counts of cited references produce 16.8% of the citation impact. The percentages for the mid-40% and bottom 50% are 51.2% and 32%, respectively. Thus, the top-10% of papers in terms of cited reference counts acquire nearly 7 percentage points more citation impact than can be expected (the expected value is 10%). In contrast, the bottom 50% in terms of cited reference counts is related to 32% of the citation impact. This is about 20% less than one can expect. Since this pattern is visible for all disciplines and publication years in Figure 3, more cited reference counts of papers seems to be related to more citations. This result is in agreement with the results of Webster, Jonason, and Schember (2009) who shows for papers in evolutionary psychology that "log citations and log references were positively related … In other words, reference counts explained 19% of the variance in citation counts" (p. 356). The results for the normalized impact groups in Figure 3 show that the normalized impact is somewhat more equally distributed over the cited references groups than the raw citation counts (see, e.g., the lower Gini coefficients for the normalized citations compared to the coefficients for times cited).

Figure 4 is based on the linked cited references which is a sub-group of all cited references (see above). The most interesting result is the very high share of citation impact ("Times cited") which is related to the 10% of papers in terms of linked cited references in the humanities: between 47.2% and 60.1% of the citation impact. These percentages are much larger than those reported in Figure 3 for all (linked and not-linked) cited references (between 24.1% and 30.5%). Thus, limiting the analysis to the linked cited references leads to an enormous increase in the inequality of citation impact and focus on the top-10% of papers in terms of cited references. However, as the right column in Figure 4 shows the normalization (of the citation impact) is able to decrease the effect of the linked cited references on the citation impact in this discipline.

Fok and Franses (2007) analyzed articles published in *Econometrica* and the *Journal of Econometrics* and found that papers with more authors tend to receive more citations. This relationship might be the "result of self-citations, but it can also be due to network effects as more authors can give more presentations at seminars and conferences and as they each may have more students who might cite their work" (Fok & Franses, 2007, p. 386). Basically, one can think of "a reference by *n* authors as having *n* times more proponents than a solo-authored one" (Valderas, 2007). The positive relationship between the number of authors and the intensity of citation impact has also been pointed out for normalized citation data (Benavent-Pérez, Gorraiz, Gumpenberger, & Moya-Anegón, 2012).

Figure 5 shows percentages of citation impact which fall upon the bottom 50% (blue segment), mid-40% (red segment), and top-10% (green segment) of papers in terms of author counts in different disciplines. For most of the disciplines in the figure, the Gini coefficients are lower than the coefficients reported in Figure 4. Thus, the citation distributions over the three author groups are more equal than the distributions over the groups based on linked cited references. However, the shares of the citation impact in Figure 5 which is related to the top-10% of papers in terms of author counts is slightly increasing in most of the disciplines (both, for impact measured in terms of non-normalized and normalized citations). The comparison of the percentages for "Times cited" and "Normalized citations" reveal that the normalization leads to a reduced effect of author counts on citation impact (especially for the humanities): The percentages based on normalized citations for the group of the top-10% papers (in terms of author counts) is lower than the percentages based on times cited. The opposite is true for the group of papers with the lowest number of authors (the papers of the bottom 50%).

In the literature, arguments can be found that the length of papers is related to the numbers of citations. For example, Gillmor (1975) reported for papers published in the *Journal of Atmospheric and Terrestrial Physics* that "the longer the paper, the more citations it received" (p. 1402). Figure 6 shows the corresponding percentages for this relationship based on all articles published in 1990, 2000, and 2010. The results point to a decreasing effect of page numbers on citation impact. For example, whereas 22% of the citation impact in the humanities account for the top-10% of papers from 1990 in terms of page numbers (see the "Times cited" column), this percentage is reduced to 13.2% in 2010. The 2010 results for nearly all disciplines ("Times cited") show distributions of citation impact over page number categories which are close to the expected values of 10%, 40%, and 50%.

The final FIC which is included in this study is the JIF, which has been studied as FIC in many other studies. The overview of Onodera and Yoshikane (2014) points out that the JIF can be considered as a strong predictor of citation impact. For example, the results of Perneger (2010) show that "the prominence of the journal where an article is published, measured by its impact factor, is positively correlated to the number of citations that the article will gather over time. Because identical articles published in different journals were compared, the characteristics of the articles themselves (be it quality of writing, scientific originality, or repute of the authors) could not have explained the observed differences. Hence, these results reflect pure journal-related bias in citation counts" (p. 662).

The benefit of receiving more citations because of a high impact of the publishing journal is debatable. Milojević, Radicchi, and Bar-Ilan (2016) argue that "receiving more citations is granted in 90% of the cases only if the IF ratio is greater than 6, while targeting a journal with an IF 2 higher than another brings in marginal citation benefit." The results of Lozano, Larivière, and Gingras (2012) further show that the citation impact of papers has increasingly been related to the JIF over the publication years. Perhaps, researchers have increasingly become aware of the importance of the JIF in evaluations and thus submit their best papers to journals with high JIF. According to Alberts (2013) JIF has even become a perverse incentive in the publication system (Leydesdorff, Wouters, & Bornmann, 2016). Given this symbolic value of JIFs in research management, it is not surprising in the science system today that researchers in some countries (e.g., developing countries) receive bonuses or salary increases when they publish in high-impact journals (Reich, 2013).

Figure 7 shows the percentage of citation impact in different disciplines which falls upon the bottom-50%, mid-40%, and top-10% when papers are ranked in terms of journal impact. In agreement with the results of Onodera and Yoshikane (2014), the JIF seems to be the strongest predictor of citation impact of single papers when compared to the other FICs. If we compare the percentages of citation impact which accounts for the top-10% of papers in terms of the different FICs in Figure 3 to Figure 7, we can mostly observe the highest values in Figure 7. The only exceptions in this comparison are the results for the humanities. For example, whereas 31% of the citation impact account for the top-10% in terms of journal impact (see Figure 7, "Times Cited, 2010"), this value is only 8.16% for page numbers, 19.7% for author counts, 20.7% for linked cited references, and 16.8% for all cited references. Similar to the results of other FICS, Figure 7 shows that the relatedness of journal impact and citation counts of papers is different for raw or normalized citations: The effect is significantly reduced if normalized citations instead of raw citation counts are used (e.g. from 31% in 2010 to 22.3% across all disciplines).

One reason for the reduction might be that the citations are normalized on the base of WoS subject categories. These categories are aggregated journal sets. Thus, the mean citation rates of the JIFs and the normalized citations might well be spuriously related.

## 5 Discussion

The skewness of citation data is frequently discussed in connection with the JIF (Bornmann & Marx, 2016; Bornmann, Marx, Gasparyan, & Kitas, 2012), although the citation data for the calculation of the JIF are less skew because they refer to a citation window of only one previous year ($t – 2$). The JIF is often used in research evaluation as a surrogate for citation scores of single papers. This practice has been frequently criticized by referring to the skewness of citations over journal papers: the results of Lariviere et al. (2016) show, for example, that "citation distributions are skewed such that most papers have fewer citations than indicated by the JIF and, crucially, that the spread of citations per paper typically spans two to three orders of magnitude resulting in a great deal of overlap in the distributions for different journals." Although this study did not investigate the skewness of citation data on the journal level, the results confirm the skewness of citation data on a much broader level, i.e., complete publication years: in 2010, nearly half of the citation impact is accounted for by the 10% most frequently cited papers, whereby the skewness is greatest in the humanities (with 68.5%) and lowest in agricultural sciences (with 40.6%). The differences, which we found between the results for the linked cited references and all cited references, further indicate that the very unequal distributions in the humanities can be traced back to the low coverage of the corresponding journals in the WoS database.

According to Seglen (1992) the skewed citation distributions can be considered as the basic probability distribution of citations which reflects "the wide range of citedness values potentially attainable and the low probability of achieving a high citation rate" (p. 632). Thus, the citation distribution seems to follow a negative binomial or Poisson distribution. This distribution is often used for situations in which many events occur with a low probability (Bornmann & Leydesdorff, 2015; Glänzel & Moed, 2013).

In statistics, it is sometimes recommended for several procedures (e.g. regression models) to exclude exceptional events occurring with low probability from the analysis (Cameron & Trivedi, 2013). In bibliometrics, however, outliers have a special meaning and many analyses specifically focus on these outliers (see, e.g., Thomson Reuters, 2016). According to Glänzel and Moed (2013), a specific issue arises in bibliometrics, "namely that of outliers. While in many fields of application outliers are simply discarded as being exceptions, in bibliometrics 'outliers' are often part of the high-end of research performance and deserve certainly special attention" (p. 382, see also Bornmann, de Moya-Anegón, & Leydesdorff, 2010).

We did not only investigate the skewness of citation distributions (in comparison with the skewness of cited references data, see Figure 2), but also the consequences of skewness in relation to covariates of citation scores. In recent years, a number of FICs have been identified potentially having an effect on citation counts. This study focused on four FICs which have been frequently investigated in the past. We used percentile shares as a method to inspect the relationship between citation impact and FICs. In our opinion, this method has the advantage that the visualization of the results allows for a straightforward interpretation of the relationship. The visualizations do not only point out the comparisons between major disciplines, but enable us to show trends over decades of research (1990, 2000, and 2010). The comparison of raw and normalized citation counts reveals whether the normalization of citations leads to a lower correlation between FICs and citation impact (this means, e.g., that the publication of a paper in a high-impact journal is less associated to more citations than its publication in a low-impact journal would have been).

The use of the term FIC in this study might suggest a causal-like relationship, not just a correlation between factors and citation impact. For instance, by interpreting the JIF as a FIC (and not just as a factor that correlates with citation impact), it is assumed that the correlation between JIF and citation impact is due to the influence (or the causal effect) of the JIF on the citation impact of an individual publication. Likewise, by interpreting the number of authors as a FIC, one assumes that a higher number of authors cause a higher citation impact. We decided to interpret the findings in this causal-like way, because the citation impact always follows the FICs on the time axis. However, these causal interpretations might be problematic if the relationship between the variables depends on a third variable (a moderator variable). For example, a third variable that may need to be taken into consideration in studies on FICs is the quality of a publication (Bornmann & Leydesdorff, 2015). If a publication is of high quality, this may cause the publication to be published in a journal with a high JIF and at the same time it may also cause the publication to receive a large number of citations. In this situation, the correlation between JIFs and citation impact is a spurious relationship and there does not need to be any causal effect of JIFs on citation impact.

The comparison of the effect of the different FICs on citation impact in this study shows (in agreement with the results of previous studies) that the JIFs have a stronger correlation than the other FICs in most of the disciplines and publication years. This is true for both raw and normalized citation counts. It is a further important result of the study for the use of citation counts for research evaluation purposes that the effect of the FICs on citation impact decreases when normalized citations instead of raw citation counts are used. It has become a standard in bibliometrics that citation counts are normalized with regard to the FICs "field", "publication year" and "document type" (but not any other FICs). Although other FICs are not considered in the normalization procedure, their effects are reduced. However, their effects do not completely vanish but remain, especially for the JIF: Whereas one can expect three times more impact for the papers from 2010 with the highest JIFs in terms of citation counts, this is reduced to two times more impact in case of using normalized citations.

One can expect that this positive effect of normalized citations is not specific to the MNCS, but will also be seen for other normalized indicators (e.g., percentiles or citing-side indicators) and/or classification schemes other than WoS subject categories. However, it would be interesting to investigate this empirically in a future study.

# Acknowledgements

The bibliometric data used in this paper are from an in-house database developed and maintained by the Max Planck Digital Library (MPDL, Munich) and derived from the Science Citation Index Expanded (SCI-E), Social Sciences Citation Index (SSCI), Arts and Humanities Citation Index (AHCI) prepared by Thomson Reuters (Philadelphia, Pennsylvania, USA).